\documentclass[page-classic]{epl2}
\shorttitle{Adding a New Dimension to DNA Melting"}

\shortauthor{S.Cuesta-Lopez et al.} 
\institute{
 \inst{1} Universit\'e de Lyon, Ecole Normale Sup\'erieure de Lyon, 
Laboratoire de Physique CNRS UMR 5672,
46 all\'ee d'Italie, 69364 Lyon Cedex 07, France\\
\inst{2} Universit\'e de Lyon, Ecole Normale Sup\'erieure de 
  Lyon, Laboratoire de Biologie Mol\'e\-culaire de la
Cellule CNRS UMR 5239,  46 all\'ee d'Italie, 69364 Lyon Cedex 7, France

}


\abstract{Standard DNA melting curves record the separation of the two strands
  versus temperature, but they do not provide any information on the location
  of the opening. We introduce an experimental method which adds a new
  dimension to the melting curves of short DNA sequences by allowing us to
  record the degree of opening in several positions along the molecule all at
  once.  This adds the spatial dimension to the melting curves and allows a
  precise investigation of the role of the base-pair sequence on the
  fluctuations and denaturation of the DNA double helix. We illustrate the
  power of the method by investigating the influence of an AT rich region on
  the fluctuations of neighboring domains.}

\begin{document}

\title{Adding a New Dimension to DNA Melting Curves.}
\author{Santiago Cuesta-Lopez$^{1}$\footnote{Mailing Address:
    Santiago.Cuesta.Lopez@ens-lyon.fr}, 
Dimitar Angelov$^{2}$, and Michel Peyrard$^1$}

\maketitle

\medskip


DNA melting, i.e.\ the separation of the two strands of the double helix, and
its reverse process hybridisation are ubiquitous in biology, in vivo for
instance for DNA transcription in the reading of the genetic code of a gene,
as well as in vitro in biological laboratories for PCR (Polymerase Chain
Reaction) or the use of DNA microarrays. This is why DNA melting has been
extensively studied even in the early days of DNA structural studies
\cite{Wartell}. An approximate understanding of the melting curves of long DNA
segments, with thousands of base pairs, can be provided by simple statistical
physics models, using empirical parameters because, at this large scale, the
subtle effects of the base pair sequence are smoothed out.  Understanding the
fluctuations and melting of short DNA fragments of a few tens of base pairs
with a high degree of heterogeneity is much more challenging. And it is also
very important because this size is the scale at which the genetic code can be
resolved. This would have some significant biological consequences to unravel
the processes by which specific binding sites are recognised by proteins,
drugs, mutagens and other molecules. This would also have a lot of practical
importance in the design of the PCR primers which are used everyday in most of
the biological laboratories \cite{Weber}.

The two kind of base pairs which exist in the DNA double helix have different
thermal stability, the AT pair, bound by two hydrogen bonds, being weaker than
the GC pair bound by three hydrogen bonds. This explains why the melting curve
of a heterogeneous DNA sequence, which shows the fraction of open pairs as a
function of temperature, can exhibit complex features. Those curves are easy
to record experimentally because the UV absorbance of a DNA solution increases
drastically when the bases are unstacked, which is the case in the broken
regions of the molecule. But such a curve only provides an integral
information on the open fraction of base pairs. Getting more local information
requires involved methods. Using a clever choice of sequences such that single
strands can form hairpins, and a combination of heating and quenching,
Montrichok et al. \cite{Montri,Montri2} managed to get some data on the melting
process of short DNA sequences, detecting whether they open at one end or by
starting with an open bubble in the centre. The kinetics of proton--deuterium
exchange for the protons involved in the hydrogen bonds within pairs, coupled
with NMR studies to detect the location of the exchanged protons, can also
provide partial information on the spatial aspect of DNA fluctuations, at the
expense of heavy experiments \cite{Dornberger}. Another approach relies on
special molecular constructs which attach a fluorophore and a quencher to DNA
to detect its local opening at a particular site \cite{BRAUNS,BONNET2003}. 
Only one
position can be monitored, and one cannot exclude local perturbations of the
fluctuations by the large residues attached to the DNA.

\medskip
Due to their importance, the statistical and dynamical properties of
DNA fluctuations and their relation to biological functions have
been the subject of many theoretical studies
\cite{Kalosakas04,us,us2}. These studies raised a
debate on the role of statistical properties and dynamical
phenomena in connection to biological function. But the validity of those
theoretical approaches can only be tested if one can compare their predictions
to measurements of the local fluctuations of the molecule. Moreover
the study of dynamical and conformational phenomena in DNA requires
a method not only able to give a precise local information, but also able to
provide coupled information of events along the chain. In this letter we
present an original method that can provide a mapping of the strength
of the fluctuations of the double helix as function of their position along
the sequence. {\it This adds a new dimension, space, to the traditional
  melting curves}.  
Our approach does not require special molecular constructs like those using
dyes or fluorophores. Instead it uses DNA itself to report on its internal
state and gives a snapshot of the opening of DNA at each guanine site at
once. 

\medskip

\begin{figure}[h]
  \centering
 \begin{tabular}{cc}
\parbox{7.5cm}{ \includegraphics[width=8cm]{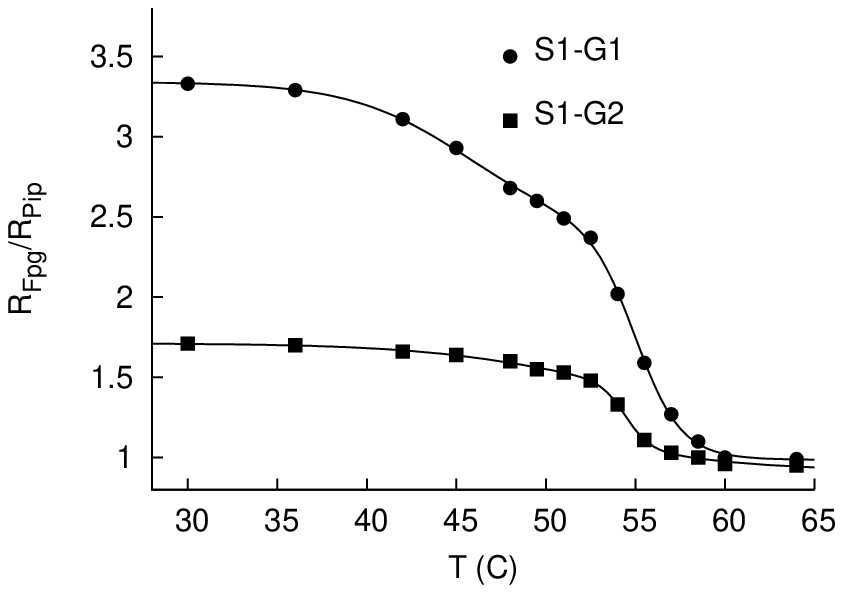} } &
\parbox{3.7cm}{
\includegraphics[width=3.6cm]{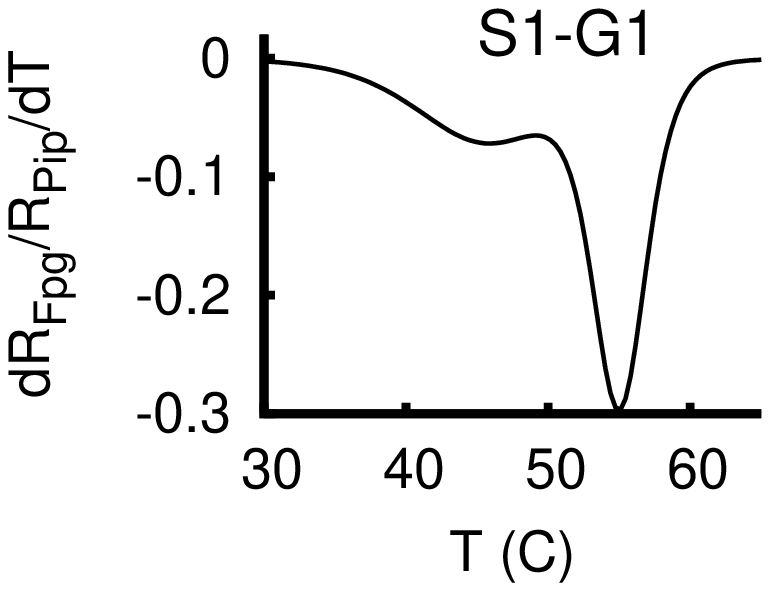}

\includegraphics[width=3.6cm]{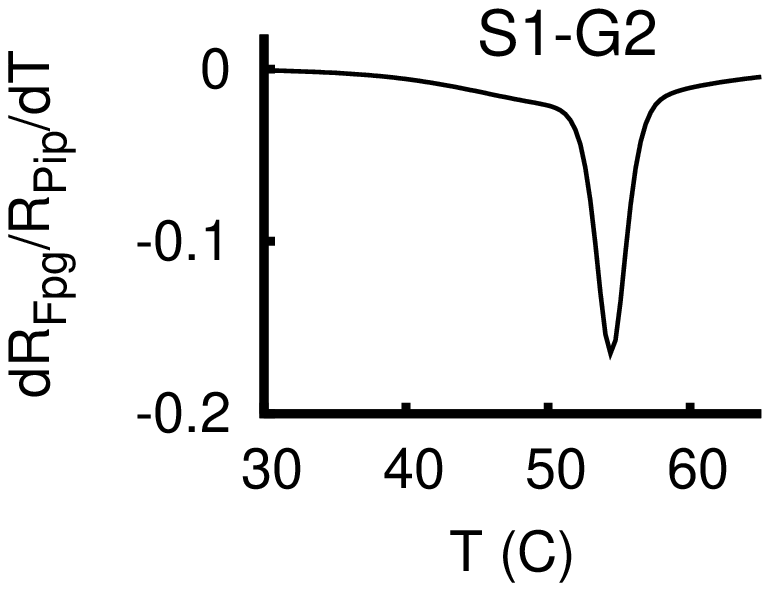}

}\\
\multicolumn{2}{c}{\includegraphics[width=10cm]{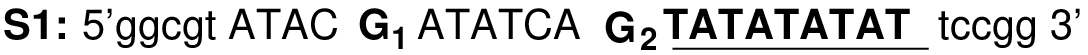}}
\end{tabular}
\caption{Local melting curves for different sections of a DNA sequence
  (S1). Symbols represent the thermal evolution of the rate $R_{Fpg}/R_{pip}$
  exhibited by two guanines $G_1$ and $G_2$, used as 
probes located along the molecule. This ratio
 reports on the closing probability of each guanine at a given temperature.
 The lines are the fitting curves described in the paper and the 
  insets plot the derivative of the fitting curve for each probe. They
  highlight the fine structure of the melting curves and 
  point out the existence of different local conformational transitions
  along the DNA sequence.}
  \label{fig:1}
\end{figure}

\begin{figure}[h]
  \centering
  \includegraphics[width=11cm]{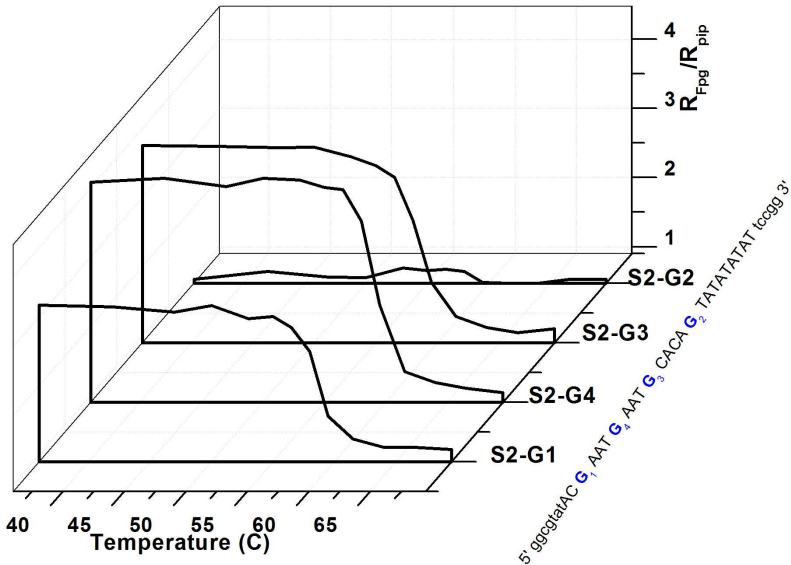}
\caption{Adding a new dimension to the study of a short DNA fragment melting
  profile: the figure displays a 3D view of the spatially resolved ratio
  $R_{Fpg}/R_{pip}$, showing the closing for different guanine probes placed
  along a longer 
  DNA sequence S2.}
  \label{fig:3D}
\end{figure}

The method relies on the oxidative chemistry of the guanine bases G, and their
propensity to be ionised by {
{a two step resonance excitation}}
\cite{Angelov05} from a strong  
UV laser pulse. Two guanine modifications, oxazolone and
8-oxo-7,8-dihydro-2-oxoguanine (8-oxodG) have been identified as the major
one-electron oxidative DNA lesions. Their formation depends on the local DNA
conformation  
and on the charge-transfer efficiency, which is affected by local
fluctuations \cite{SPASSKY2002,Douki,Spassky97}. While oxazolone is the unique
product resulting from one-electron oxidation of the free 2$'$-deoxyguanosine,
8-oxodG appears as soon as the nucleoside is incorporated in a helical
structure. Hence the measurement of the relative yield of these photoproducts
at each G site (labelled $R_{Fpg}/R_{pip}$ due to the
method used for its detection \cite{note1}) 
tells us whether this G was in an helical structure (closed) or
whether it was open when the
molecule was hit by the laser pulse. As the experiment is not performed on a
single molecule but in solution the results are obtained on a statistical
ensemble and they give a signal representative of the probability that
each G site is closed at the temperature of the study. 
Standard biological methods
can be used to measure the relative yield of the
production of oxazolone and 8-oxodG \cite{note1}.
However, it is important to notice that the value of $R_{Fpg}/R_{pip}$  
should not be
considered as a quantitative measure of the local closing probability
because it is also affected by the configuration of the DNA molecule near the
probe, which depends slightly on the sequence and influence the charge
transfer. Only the temperature dependence of this ratio for a given probe can
be analysed quantitatively \cite{note1b}.

\smallskip
By splitting the sample into several aliquots, the measurement can be
performed at different temperatures, which allows us to produce a set of
melting curves for each guanine in the sequence. Figure 1 shows the results of
a UV laser irradiation analysis for two guanines, labelled S1-G1 and S1-G2,
belonging to a test sequence {\bf S1} (details of the sequence are provided in
the figure caption and in Figure 3). It demonstrates how our method reports
two complementary curves at once for {{
 {the}} same single DNA
  sequence, adding a valuable information of the state of the system. The
  results that it provides changes the view in which DNA denaturation can be
  studied, adding the spatial correlations to the notion of local conformation
  (melted or packed helix).

In order to quantitatively analyse the measurements shown in Fig.~1 
{
{we have fitted the experimental curves by the function $f(T) = A - B_1
  \tanh[C_1(T-T_1)] - B_2 \tanh[C_2(T-T_2)]$, selected according to the shape
  of the curves, particularly that of the probe S1-G1. Once the optimal
  parameters are determined, we plot $df(T)/dT$ for each probe (right part of
  Fig.~\ref{fig:1}),}} {
{to highlight the fine structure of each
    melting curve. Figure~\ref{fig:1} clearly shows that, at the level of
    probe S1-G1, the melting occurs in two steps, with a precursor at $T_2 =
    45.7^{\circ}$C while the full melting is achieved at
    $55.0^{\circ}$C. Although a slight precursor effect can be detected for
    probe S1-G2 it is very weak and hardly visible on the figure. The fit of
    the data indicates that it occurs at $T'_2=51.8^{\circ}$C, very close to
    the full melting detected at $T'_1=54.4^{\circ}$C for this
    probe. Interestingly these results show that the conformational melting
    can follow different paths in various parts of the same short sequence.}}

For longer sequences such as sequence {\bf S2} (Fig.~\ref{fig:Sequences}), 
the existence of several guanine probes, located in some interesting domains,
allows us to obtain a collection of snapshots for the local state of the
system that can be combined to build a 
three-dimensional melting profile like the one shown in Figure 2. This 3D plot
shows the ratio $R_{Fpg}/R_{pip}$ for a particular DNA sequence containing
four guanines (S2-G1 to S2-G4) that are monitored as probes. 
The study as a function of temperature provides a melting profile for each
probe, together with a view 
of the spatial correlations along the full sequence.

\begin{figure}[h]
  \centering
  \includegraphics[width=11cm]{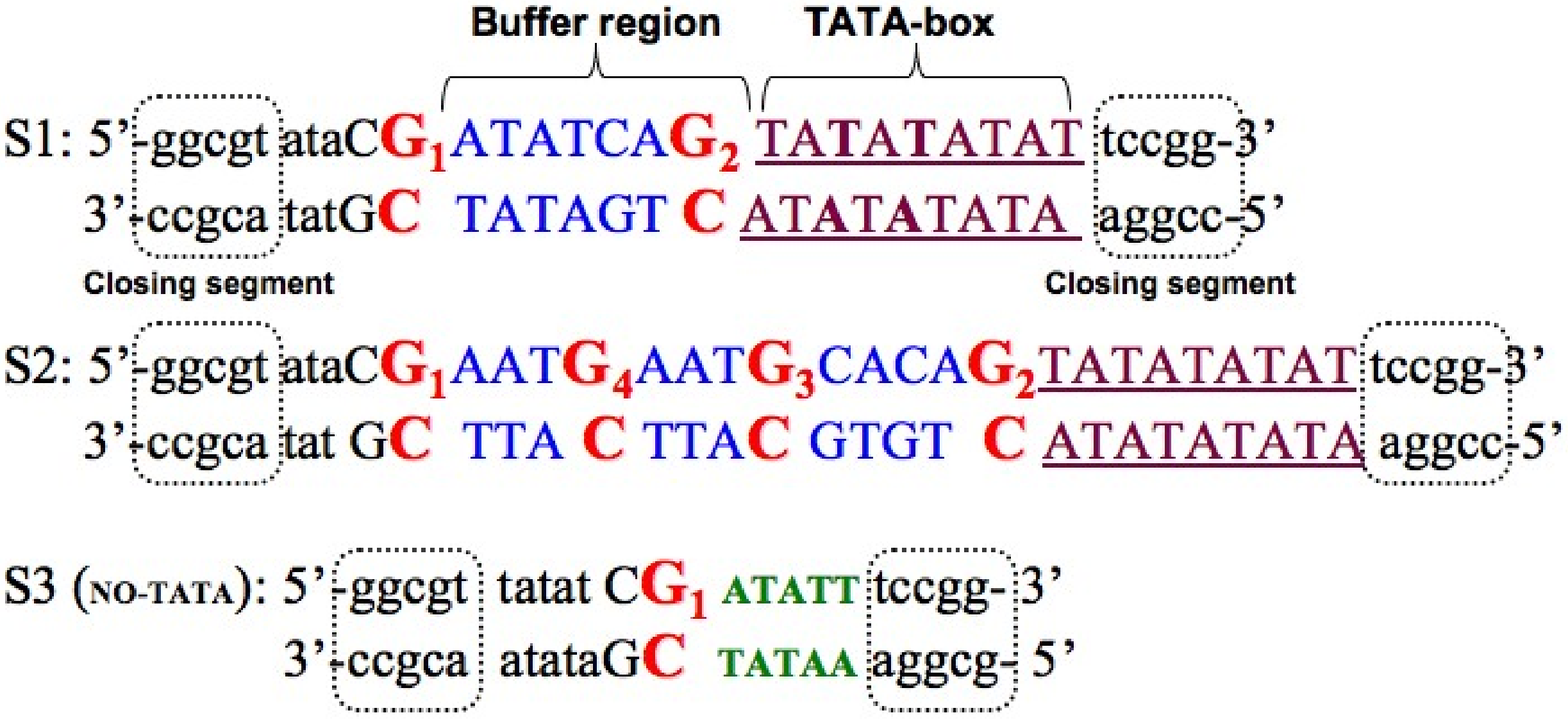}
\caption{Sequences of the DNA fragments investigated in this study. {\bf S1}
  and {\bf S2}, are artificial sequences containing a large TATA box and
  guanines on the 5'-3' strand, used as probes. {\bf S3} is a shorter {\em
    control} sequence that eliminates the TATA motif. All sequences are
  completed by GC-rich terminal domains (marked as dotted boxes) to stabilise
  them and ensure proper closing of these short DNA helices.}
  \label{fig:Sequences}
\end{figure}

In order to get a more precise insight on the origin of the two-step melting
detected at the level of probe S1-G1, and to relate it to the effect of bubble
nucleation in DNA, we have studied several artificial sequences specially
designed and synthesised \cite{note2} to investigate possible non-local
effects of the fluctuations.  They are shown on Figure 3. Sequences {\bf S1} 
and {\bf S2}
contain a 10 base-pair-long AT-rich fragment which is analogous to the
``TATA-box', 
a motif that exists in the transcription-initiation regions of the
genes of various species. As the AT base pairs bound by only two hydrogen
bonds are weaker than the GC pairs, these segments are expected to exhibit
large fluctuations even at biological temperature because they are closer to
their melting temperature.

All sequences {\bf S1}, {\bf S2} and {\bf S3} 
contain different guanines labeled Sn-Gx,
where $n$ refers to the sequence and $x$ to the particular guanine.
They are spread along the strand, and according to the basis of our method,
act as 
probes informing us about the local structural state at each temperature. 
As mentioned above the level of the $R_{Fpg}/R_{pip}$ signal that we record
depends on the structure in the vicinity of the guanine of interest.
This is why, in order to allow a quantitative comparison between different
sequences, we have selected guanines with the same environment. For instance
guanines S1-G1, S2-G1, S3-G1 are all parts of the sequence CGA,
and guanines S1-G2 and S2-G2 are part of the sequence AGT.
The major difference between sequences {\bf S1} and {\bf S2} 
lies in the length of the domain that separates probe G1 from the AT-rich
region. This ``buffer 
region'', which is an heterogeneous domain with AT and GC pairs, 
has been extended from 7 base pairs in sequence {\bf S1} to 13 base pairs
in sequence {\bf S2}. In sequence {\bf S3} 
we have eliminated the large ``TATA box'' to
keep only short AT-rich domains around probe S3-G1. In all cases these short
DNA molecules are terminated by GC rich domains which act as clamps to prevent
large fluctuations of the free ends of 
the molecules, and hold the two strands together even when we heat the sample
up to $60^{\circ}$C. \cite {note3}.

The variation versus temperature of the ratios $R_{Fpg}/R_{pip}$ for all the
guanine probes studied in sequences {\bf S1}, {\bf S2}, {\bf S3}, 
are summarised in
Fig.~\ref{fig:FitsS2S3}. The comparison of the various curves gives some
clues to understand the differences between the two melting curves of probes
S1-G1 and S1-G2 discussed above, and points out some interesting features of
DNA fluctuations, which can be revealed by an experimental method able to
record local melting profiles.
When they are analysed in the
context of the particular sequences that we studied, the curves suggest
three important properties of DNA fluctuations:\smallskip

i) a large AT-rich domain undergoes very large fluctuations, even at room or
biological temperature, and therefore tends to easily form an ``open bubble''.

ii) there is a minimum size of the AT-rich domain that allows the formation of
such a bubble.

iii) the influence of such a bubble does not only affect its immediate
vicinity, but extends to some distance.

Let us see how these statements are supported by our results.

As our method relies on the ionisation of the guanines, we do not directly
measure the opening of the AT pairs but their fluctuations can be
inferred from their influence on the adjacent guanines.  Although we stressed
that quantitative comparisons cannot be made between different probes because
the signal that we record depends on the local structure of DNA, the very small
value of $R_{Fpg}/R_{pip}$ for probe S1-G2 is nevertheless
a strong indication that the
closing probability of this guanine, which lies next to a series of 10 AT
pairs is very low even at room temperature. 
This can be understood as an effect of the strong
tendency of the large AT rich region to open into transient bubbles, called
``premelting phenomena'', starting at physiological temperatures
\cite{ERFURTH,CHAN90,CHAN93,MOVILEANU2002}, which certainly affects the base
pair which is right next to it. Probe S3-G1 is also surrounded by AT-rich
regions, but its closing probability deduced from the corresponding value of
$R_{Fpg}/R_{pip}$, shown on the top panel of Fig.~\ref{fig:FitsS2S3},
appears to be much higher that for probe S1-G2. This indicates that the
fluctuations of the five-base-long AT regions which are next to probe S3-G1
are not sufficient to form open bubbles that would promote the opening of
this probe.  This is in agreement with the existence of a minimum size needed
to allow the formation of a bubble \cite {Montri}.

The striking point is that the large fluctuations of the TATA box
 do not only perturb the adjacent geometry but also induce conformational
 changes that 
distort the closed packed helicoidal structure in regions distant from the
bubble nucleation segment, giving rise to {\it premelting intermediate
  structural states} that coexist in consonance with the nucleated
bubbles. This shows up in the two-step denaturation that we observed for probe
S1-G1, as discussed above. Although this probe is 7 base-pairs away 
from the TATA box, Fig.~\ref{fig:1} shows that precursor effects appear well
below the full denaturation of this probe. Those precursors are
{
{also visible in Fig.~\ref{fig:FitsS2S3}. Note again how the
    melting at the level of probe S1-G1 differs from a simple sigmoidal curve,
    particularly in the temperature range from 37$^{\circ}$C to
    52$^{\circ}$C.}} 
It is tempting to assign them to the influence of the fluctuations of the
TATA, which grow when the temperature is increased and might influence the
opening structure and fluctuations of the double helix even rather far
away. But, to confirm such an assignment control experiments are
necessary. Their results are shown on the {
 {bottom}} panel of
Fig.~\ref{fig:FitsS2S3}  which shows the thermal variation of the
ratio $R_{Fpg}/R_{pip}$ for several guanine probes in sequence {\bf S2}.

In this sequence the TATA box is present, as in sequence {\bf S1}, 
but the length of the buffer region that separates it from probe S2-G1, which
has the same local 
structure as probe S1-G1, has been increased. Moreover this buffer region is
strengthened because it contains several GC pairs, including two adjacent ones
at the site of probe S2-G3. In this sequence the existence of the large
fluctuations of the TATA box are attested by the very low closing probability
of probe S2-G2, similar to what is observed for probe S1-G2. But, contrary to
what was observed for probe S1-G1, the denaturation of probe S2-G1 does not
show any {
{significant}} precursor effect. Similar one-step
transitions are also observed for probes S2-G3 and S2-G4.

In summary, these experiments clearly suggest that thermal fluctuations, which
are stronger in AT-rich tracks, induce bubble nucleation phenomena and
structural changes that affect not only the local geometry and dynamics of DNA
in the breathing portion, but also to some distance along the helix. 
How this happens is highly dominated by the characteristics of the fragment
sequence as well as the length of the sections involved.\

\begin{figure}[h!]
  \begin{center}
\includegraphics[width=8.6cm]{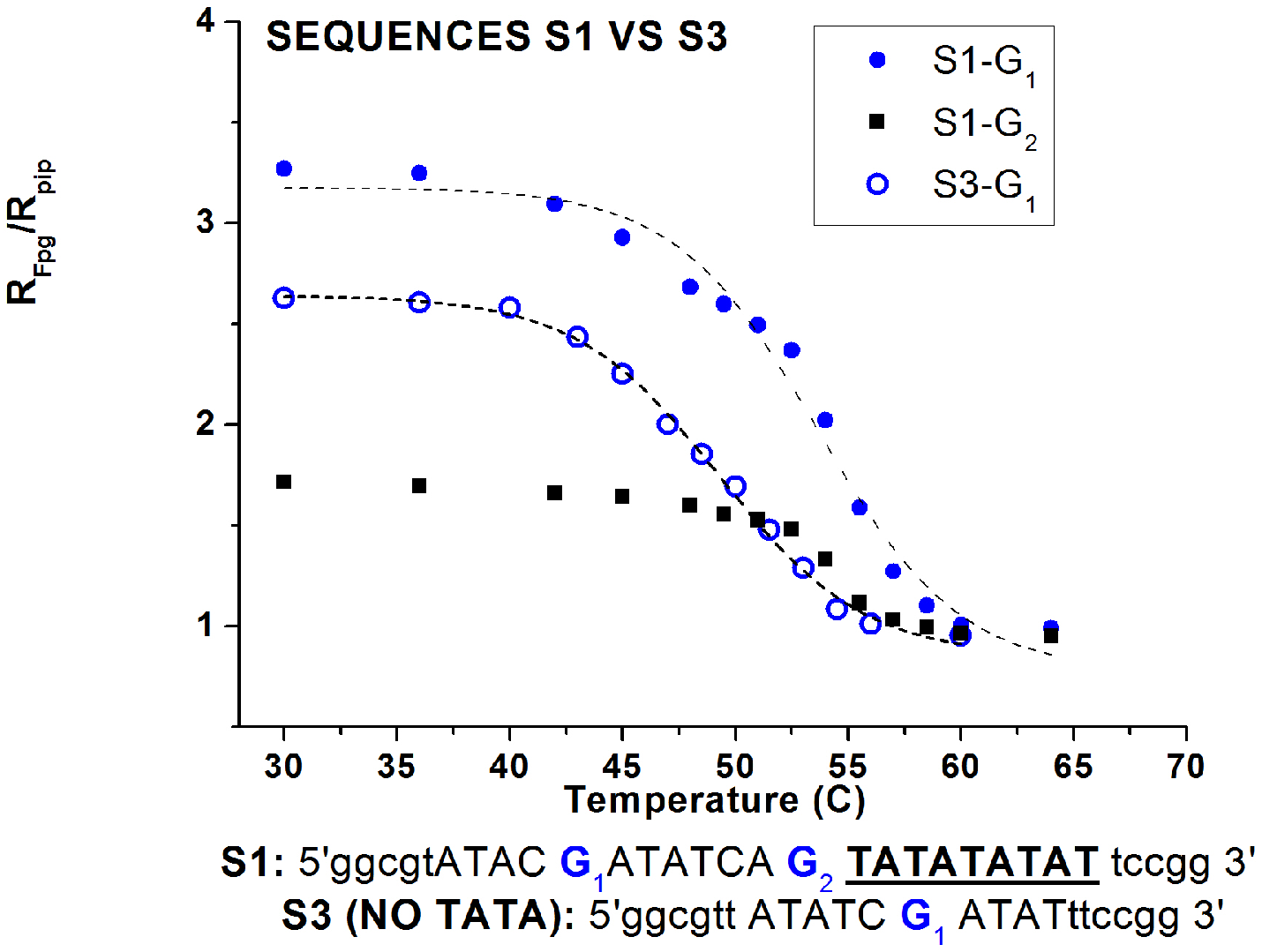}
\hspace{-1.148cm}
\raisebox{0.25cm}{
\includegraphics[width=8.6cm]{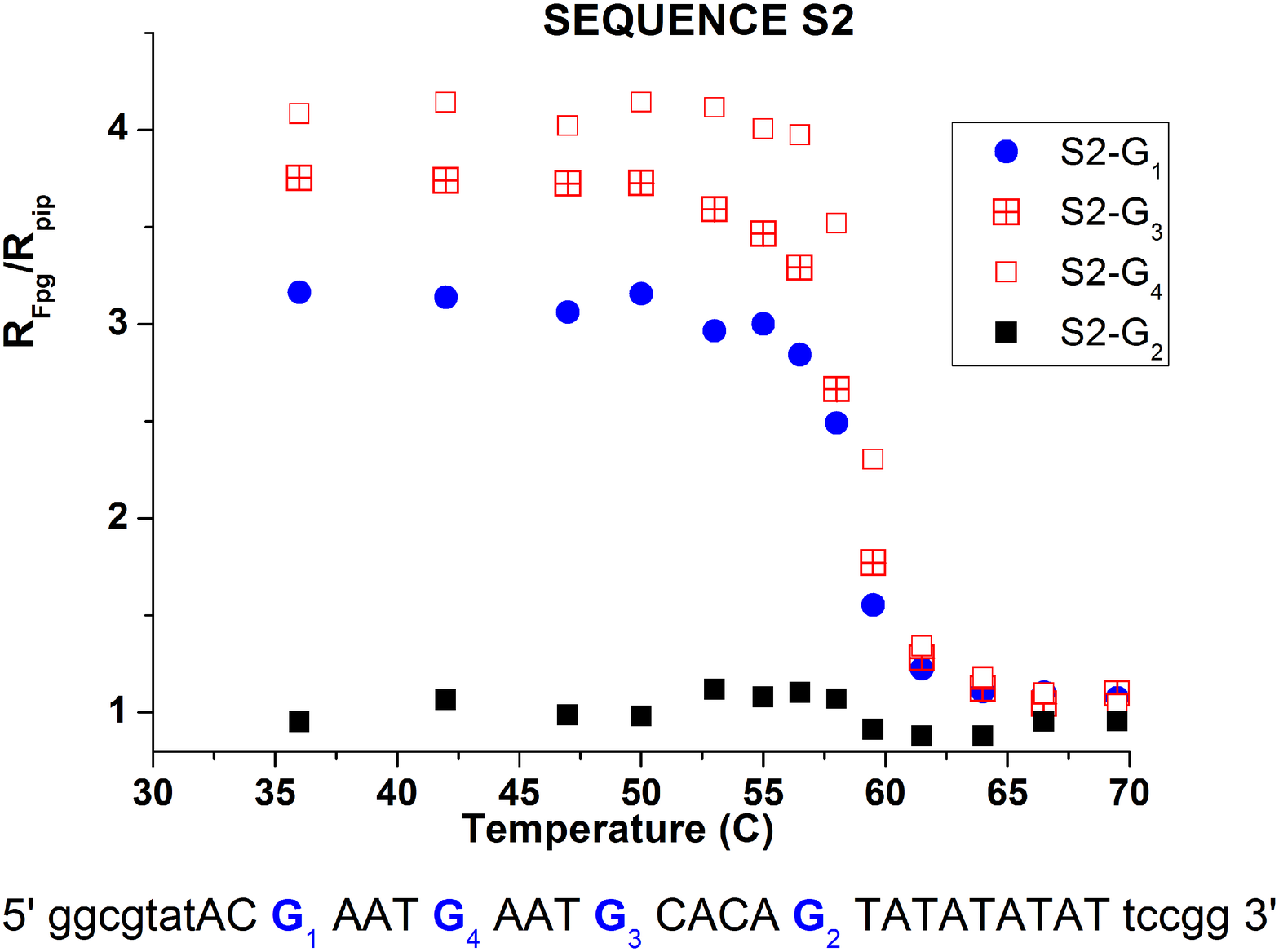}
} 
  \end{center}
\caption{The top panel shows the melting profiles for the probes studied in
  sequences {\bf S1} and {\bf S3}. Sigmoidal non-linear fits (dotted lines)
  have been used to describe the evolution of probes S1-G1 and S3-G1. Both
  fits have been performed with the same identical functional form. While the
  response of S3-G1 agrees with a one step standard melting transition, the
  result for S1-G1 suggests a two steps transition, with a {\it premelting}
  region,  clearly shown in Fig.~\ref{fig:1}. The bottom panel shows the ratio
  $R_{Fpg}/R_{pip}$ for the guanine probes of sequence {\bf S2} versus
  temperature. For probe S2-G1, the premelting effect has been suppressed due
  to the existence of a longer more stable intermediate buffer region.} 
\label{fig:FitsS2S3}
\end{figure}

Those results are only accessible because we have introduced a method which is
able to provide a spatial information that was not available until now for the
study of local melting in short DNA fragments. Further studies are certainly
necessary to confirm 
and precise the non-local effect of the fluctuations of large AT-rich regions
that we have presented in 
this letter. They become possible with the ``three-dimensional melting
curves'' that can now be measured. 

Finally, we would like to emphasise the importance of taking into account the
structural modifications induced by bubble dynamics in terms of DNA-protein
binding interactions, Transcription Factor recognition or DNA-drug
binding. Further studies of local fluctuations in DNA may be of significant
importance in the analysis of these biological phenomena. 

\smallskip
We would like to thank the program CIBLE of R\'egion Rh\^one--Alpes which
supported this work.


\begin{thebibliography}{9}
\bibitem{note1}
To measure the relative yield of the production of oxazolone and 8-oxodG we
use common chemical methods used in standard analytical biology: Piperidine
cleaves DNA at the sites where oxazolone is the resulting lesion whereas Fpg
protein cleaves it as sites which have evolved to 8-oxodG.  One end of the
molecule is marked by a radioactive marker, and, after cleavage by piperidine
or Fpg, performed on different aliquots at the same temperature, the lengths
of the fragments that carry the marker are quantified by gel
electrophoresis. The radioactively 
 labelled fragments of length $l$ are those which were
cleaved at a guanine situated at distance $l$ from the marked end. The ratio
$R_{Fpg}$/$R_{pip}$ of the number of radioactively labeled fragments of length
$l$ produced by Fpg cleavage or by piperidine cleavage indicates the
probability of closing for this particular guanine{
, which is
  unambiguously identified by its distance from the marked end}. As the gel
image contains 
the data for fragments of all lengths, a single experiment measures a
magnitude proportional to the opening probability at all guanine sites.
\bibitem{note1b}
The experimental errors are within the size of the data points shown in all
graphs.
\bibitem{note3} We should point out 
  that the spatial resolution near the end of the
  strands in the gel is usually not optimal to separate and resolve two
  adjacent guanines in this region. In principle the guanines in these
  terminal parts could also be monitored in our analysis, but such limitation
  makes the study inefficient.
\end{thebibliography}
\end{document}